\documentclass{ws-procs975x65}

\begin{document}



\title{KERR'S GRAVITY AS A QUANTUM GRAVITY ON THE COMPTON
LEVEL\footnote{Talk at the QG1 session of the MG11 meeting,
partially supported by RFBR grant 07-08-00234.}}

\author{ALEXANDER BURINSKII
}

\address{Gravity Research Group, NSI Russian Academy of Sciences,\\
B. Tulskaya 52, Moscow 115191, Russia, \email{bur@ibrae.ac.ru}}


\def\b{\bar}
\def\d{\partial}
\def\D{\Delta}
\def\cA{{\cal A}}
\def\cD{{\cal D}}
\def\cK{{\cal K}}
\def\cF{{\cal F}}
\def\f{\varphi}
\def\g{\gamma}
\def\G{\Gamma}
\def\l{\lambda}
\def\L{\Lambda}
\def\M{\mathcal{M}}
\def\m{\mu}
\def\n{\nu}
\def\p{\psi}
\def\q{\b q}
\def\r{\rho}
\def\t{\tau}
\def\x{\phi}
\def\X{\~\xi}
\def\~{\tilde}
\def\h{\eta}
\def\bZ{\bar Z}
\def\cY{\bar Y}
\def\bY3{\bar Y_{,3}}
\def\Y3{Y_{,3}}
\def\z{\zeta}
\def\Z{{\b\zeta}}
\def\Y{{\bar Y}}
\def\cZ{{\bar Z}}
\def\`{\dot}
\def\be{\begin{equation}}
\def\ee{\end{equation}}
\def\bea{\begin{eqnarray}}
\def\eea{\end{eqnarray}}
\def\half{\frac{1}{2}}
\def\fn{\footnote}
\def\bh{black hole \ }
\def\cL{{\cal L}}
\def\cH{{\cal H}}
\def\cP{{\cal P}}
\def\cM{{\cal M}}
\def\ol{\overline}
\def\const{{\rm const.\ }}
\def\ik{ik}
\def\mn{{\mu\nu}}
\def\a{\alpha}

\begin{abstract}
The Dirac theory of electron and QED neglect gravitational field,
while the corresponding to electron  Kerr-Newman
gravitational field has very strong influence on the Compton
distances. It polarizes space-time, deforms the Coulomb field and
changes topology. We argue that the Kerr geometry may be hidden
beyond the Quantum Theory, representing a complimentary space-time
description.
\end{abstract}

\bodymatter

{\bf 1.}{\bf Introduction.} The Kerr-Newman solution displays many
relationships to the quantum world. It is the anomalous
gyromagnetic ratio $g=2$,  stringy structures  and other features
allowing one to construct a semiclassical model of the extended
electron \cite{Bur0,BurTwi,BurAxi,Rengra} which has the Compton
size and possesses the wave properties. Meanwhile, the quantum
theory neglects the gravitation at all. The attempts to take into
account gravity are undertaken by superstring theory which is
based on the space-time description of the extended stringy
elementary states: $\quad Points \quad \longrightarrow \quad
Extended \ Strings,$ and also, on the unification of the Quantum
Theory with Gravity on Planckian level of masses $ M_{pl},$ which
correspond to the distances of order $10^{-33}$ cm.

Note, that spin of quantum particles is  very high with respect to
the masses. In particular, for electron $S= 1/2 ,$ while $
m\approx 10^{-22}$ (in the units $G=\hbar=c=1$). So, to estimate
gravitational field of spinning particle, one has to use the Kerr,
or Kerr-Newman solutions \cite{DKS}, contrary to the ordinary
estimates based on spherical symmetric solutions.

{\it Performing such estimation, we obtain a striking
contradiction with the above scale of Quantum Gravity !}

Indeed, for the Kerr and Kerr-Newman solutions we have the basic
relation between angular momentum $J$, mass $m$ and radius of the
Kerr singular ring $a$ : $ J=ma .$ Therefore,
Kerr's gravitational field of  a spinning particle is extended
together with the Kerr singular ring up to the distances $a= J/m
=\hbar /2m \sim  10^{22}$ which are of the order of the Compton
length of electron $10^{-11}$ cm., forming a singular closed
string\fn{See also \cite{Isr,Bur0,Lop,BurOri}.}. Since $a>>m ,$
this string is naked (no event horizon of black hole).
In the Kerr geometry, in
analogy with string theory {\it the `point-like' Schwarzschild
singularity turns into an extended  string of the Compton size. }

Note, that the Kerr string is not only analogy. It was shown
that the Kerr singular ring is indeed the string \cite{BurOri},
and, in the analog of the Kerr solution to low energy string
theory \cite{Sen}, the field around the Kerr string is similar to
the field around a heterotic string \cite{BurSen}.  It is an
Alice topological string \cite{BurTwi,Rengra}, and the Kerr space
exhibits a change of topology on the Compton distances.
Therefore, the Kerr geometry indicates  essential peculiarities of
space-time on the Compton distances, and the
use of Kerr geometry for estimation of the scale of Quantum
Gravity gives the striking discrepancy with respect to the
ordinary estimations based on the Schwarzschild geometry.

{\it There appears the Question: ``Why  Quantum Theory does not
feel such drastic changes in the structure of space time on the
Compton distances?''}
How can such drastic changes in the structure of  space-time and
electromagnetic field  be experimentally unobservable and
theoretically ignorable in QED?

There is, apparently, unique explanation to this contradiction.  We
have to assume that the Kerr geometry {\it is already taken into
account in quantum theory} and play there an important role. In
another words, the Kerr geometry is a complimentary (dual)
space-time description of quantum processes.

\begin{figure}[ht]
\centerline{\epsfig{figure=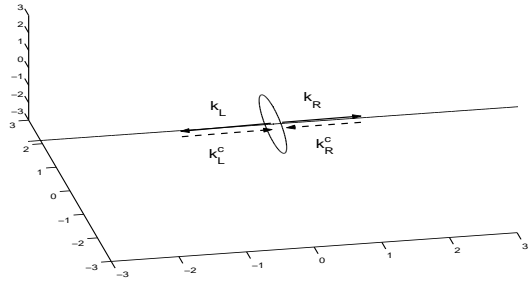,height=4cm,width=7cm}}
\caption{Skeleton of the Kerr spinning particle in the
rest frame: the Kerr singular ring and two semi-infinite singular half-strings
which are determined by two null-vectors of polarization of a
free electron.} \end{figure}

Indeed, the local gravitational field at these distances is
extremely small, for exclusion of an extremely narrow vicinity of
the Kerr singular ring forming a closed string of the Compton
radius. This closed Kerr string is presumably the source of
quantum effects.

Such point of view coincides with the old conjecture on the Kerr
spinning particle as a model of electron, a
`microgeon' model, where the spin and mass of electron are related
with e.m. and spinor excitations of the Kerr closed string
\cite{Bur0,BurTwi,BurAxi}.
The compatible with the Kerr geometry`aligned' excitations
\cite{BurTwi,BurAxi} have a peculiarity in the form of two extra
semi-infinite singular half-strings, as it is shown on fig.1.

Excitations of the Kerr circular string of the Compton size
$a=\hbar/m$ have the wave lengths $\lambda =\frac a {2n},$
and, as usual in string theory, generate the mass
$m=E=\hbar c/\lambda$ and spin of particle $J=ma = \hbar/2 .$
In the same time, the waves induced by excitations on the axial
strings carry de'Broglie periodicity \cite{BurAxi,BurTwi}.

\begin{figure}[ht]
\centerline{\epsfig{figure=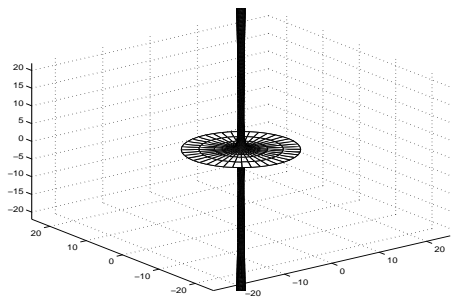,height=4cm,width=6cm}}
\caption{Image of the dressed Kerr spinning particle.}
\end{figure}

Vacuum polarization near the singular strings leads to
the formation of a false vacuum, so there has to be a phase
transition near the sources \cite{Rengra}, and the Kerr spinning
particle turns out to be dressed, taking the form shown on fig.2.

One of the often discussed objections against the Compton size of
electron is the argument based on the experiments on the deep inelastic
scattering of electron which demonstrates its almost point-like
structure. Explanation of this fact may be divided onto two parts:

a) the point like exhibition
of the structure of electron may be related with the complex
representation of the Kerr source which {\it is point-like from
the complex point of view} \cite{BurTwi,Beyond}.
Working in the momentum space,
one can feel namely this point-like structure. On the real
space-time slice it is realized as a contact interaction of the
`axial' strings \cite{BurTwi};

b) the space-time Compton extension of electron has also been
observed in the low-energy experiments with a coherent resonance
scattering of electron\cite{BLP}. In this relation,
the experiments with polarized electrons has to be the most
informative.

Finally, one can mention the obtained recently multiparticle
Kerr-Schild solutions \cite{Multiks} which show that theory
of electron is to be multiparticle one, indeed.

\vfill

\end{document}